\documentclass[journal,twocolumn]{IEEEtran}
\usepackage{amsmath,graphicx}
\usepackage{color}
\usepackage{graphicx}
\usepackage{epstopdf}
\usepackage{amsmath}
\usepackage{amssymb}
\usepackage{mathrsfs}
\usepackage{hyperref}
\usepackage{tikz}
\usepackage{bm}
\usepackage[english]{babel}
\usepackage{cite}
\usepackage{rotfloat}
\usepackage{mathtools}
\usepackage[font=normalsize,labelfont=bf]{caption}
\usepackage{amsmath}
\usepackage{makecell}
\usepackage{algorithm,algorithmic}
\usepackage{multirow}
\usepackage{subfigure}
\usepackage{booktabs}
\usepackage{colortbl}
\usepackage{multirow}
\usepackage{hhline}
\usepackage{stfloats}
\usepackage{multicol}
\usepackage{bbm}
\usepackage{cases}
\graphicspath{ {Figures/} }
\newsavebox{\foobox}

\definecolor{kugray5}{RGB}{224,224,224}

\usepackage[normalem]{ulem}
\newcommand\rsout{\bgroup\markoverwith
	{\textcolor{red}{\rule[0.5ex]{2pt}{0.8pt}}}\ULon}



\makeatletter
\newcommand{\ALOOP}[1]{\ALC@it\algorithmicloop\ #1%
	\begin{ALC@loop}}
	\newcommand{\ENDALOOP}{\end{ALC@loop}\ALC@it\algorithmicendloop}

\makeatother

\usepackage{etoolbox}
\let\mybibitem\bibitem
\renewcommand{\bibitem}[1]{%
	\ifstrequal{#1}{nature}
	{\color{blue}\mybibitem{#1}}
	{\color{black}\mybibitem{#1}}%
}

\graphicspath{ {Figures/} }

\DeclareCaptionLabelSeparator{periodspace}{.\quad}

\renewcommand{\figurename}{Fig.}
\addto\captionsenglish{\renewcommand{\figurename}{Fig.}}
\allowdisplaybreaks

\usepackage{setspace}

\def\b0{{\pmb{0}}}

\usepackage{caption}
\usepackage{xspace}

\usepackage{acronym}
\acrodef{cs}[C\&S]{\emph{communications and sensing}}
\acrodef{ai}[AI]{\emph{artificial intelligence}}
\acrodef{ml}[ML]{\emph{machine learning}}
\acrodef{csi}[CSI]{{channel state information}}
\acrodef{mimo}[MIMO]{multiple-input-multiple-output}
\acrodef{mmwave}[mmWave]{millimeter-wave}
\acrodef{aps}[APs]{access points}
\acrodef{cpu}[CPU]{central processing unit}
\acrodef{los}[LoS]{line-of-sight}
\acrodef{nlos}[NLoS]{non-line-of-sight}
\acrodef{wp}[WP]{work package}
\acrodef{isac}[ISAC]{\emph{integrated sensing and communications}}
\acrodef{ris}[RIS]{reconfigurable intelligent surface}
\acrodef{cl}[Cooperative Learning]{}
\acrodef{scnr}[SCNR]{signal-to-cluster-plus-noise ratio}
\acrodef{kasit}[KAIST]{Korea Advanced Institute of Science and Technology}
\acrodef{asu}[ASU]{Arizona State University}
\acrodef{cwc}[CWC-UOULU]{Centre for Wireless Communications, University of Oulu}
\acrodef{uoulu}[UOULU]{University of Oulu}
\acrodef{rcf}[RCF]{Research Council of Finland}
\acrodef{crlb}[CRLB]{Cramér–Rao lower bound}
\acrodef{sinr}[SINR]{signal-to-interference-plus-noise ratio}

\usepackage[table]{xcolor}
\usepackage{pgfmath}

\usepackage{nicematrix}
\usepackage{xcolor}

\definecolor{fullblack}{gray}{0.0}    
\definecolor{darkgray}{gray}{0.3}    
\definecolor{mediumgray}{gray}{0.4}   
\definecolor{lightgray}{gray}{0.48}    
\newcommand{\colorvalImprovement}[1]{%
  \ifdim #1 pt > 4pt
    \textcolor{fullblack}{\textbf{#1}}%
  \else\ifdim #1 pt > 2.9pt
    \textcolor{darkgray}{\textbf{#1}}%
  \else\ifdim #1 pt > -3pt
    \textcolor{mediumgray}{#1}%
  \else\ifdim #1 pt > -5pt
    \textcolor{darkgray}{#1}%
  \else
    \textcolor{darkgray}{\textbf{#1}}%
  \fi\fi\fi\fi
}
\newcommand{\colorvalLoss}[1]{%
  \ifdim #1 pt < -5pt
    \textcolor{fullblack}{\textbf{#1}}%
  \else\ifdim #1 pt < -1pt
    \textcolor{fullblack}{\textbf{#1}}%
  \else\ifdim #1 pt < 3pt
    \textcolor{mediumgray}{#1}%
  \else\ifdim #1 pt < 5pt
    \textcolor{lightgray}{{#1}}%
  \else
    \textcolor{lightgray}{{#1}}%
  \fi\fi\fi\fi
}

\usepackage[all=normal,paragraphs=tight,floats=normal,mathspacing=normal,wordspacing=tight,charwidths=tight,mathdisplays=normal,leading=normal]{savetrees}
\usepackage{soul}

\begin{document}
\title{Knowledge Distillation for Collaborative Learning \\in Distributed Communications and Sensing}
\author{Nhan~Thanh~Nguyen,
        Mengyuan~Ma,
        Nir~Shlezinger,
        Junil~Choi,\\
        Yonina~C.~Eldar,
        A.~Lee~Swindlehurst,
        and Markku~Juntti
}

\maketitle

\begin{abstract}
The rise of sixth generation (6G) wireless networks promises to deliver ultra-reliable, low-latency, and energy-efficient communications, sensing, and computing. However, traditional centralized artificial intelligence (AI) paradigms are ill-suited to the decentralized, resource-constrained, and dynamic nature of 6G ecosystems. This paper explores knowledge distillation (KD) and collaborative learning as promising techniques that enable the efficient and scalable deployment of lightweight AI models across distributed communications and sensing (C\&S) nodes. We begin by providing an overview of KD and highlight the key strengths that make it particularly effective in distributed scenarios characterized by device heterogeneity, task diversity, and constrained resources. We then examine its role in fostering collective intelligence through collaborative learning between the central and distributed nodes via various knowledge distilling and deployment strategies. Finally, we present a systematic numerical study demonstrating that KD-empowered collaborative learning can effectively support lightweight AI models for multi-modal sensing-assisted beam tracking applications with substantial performance gains and complexity reduction.
\end{abstract}\vspace{-0.25cm}

\section{Introduction}

\begin{figure}
    \centering
    \includegraphics[width=1.0\linewidth]{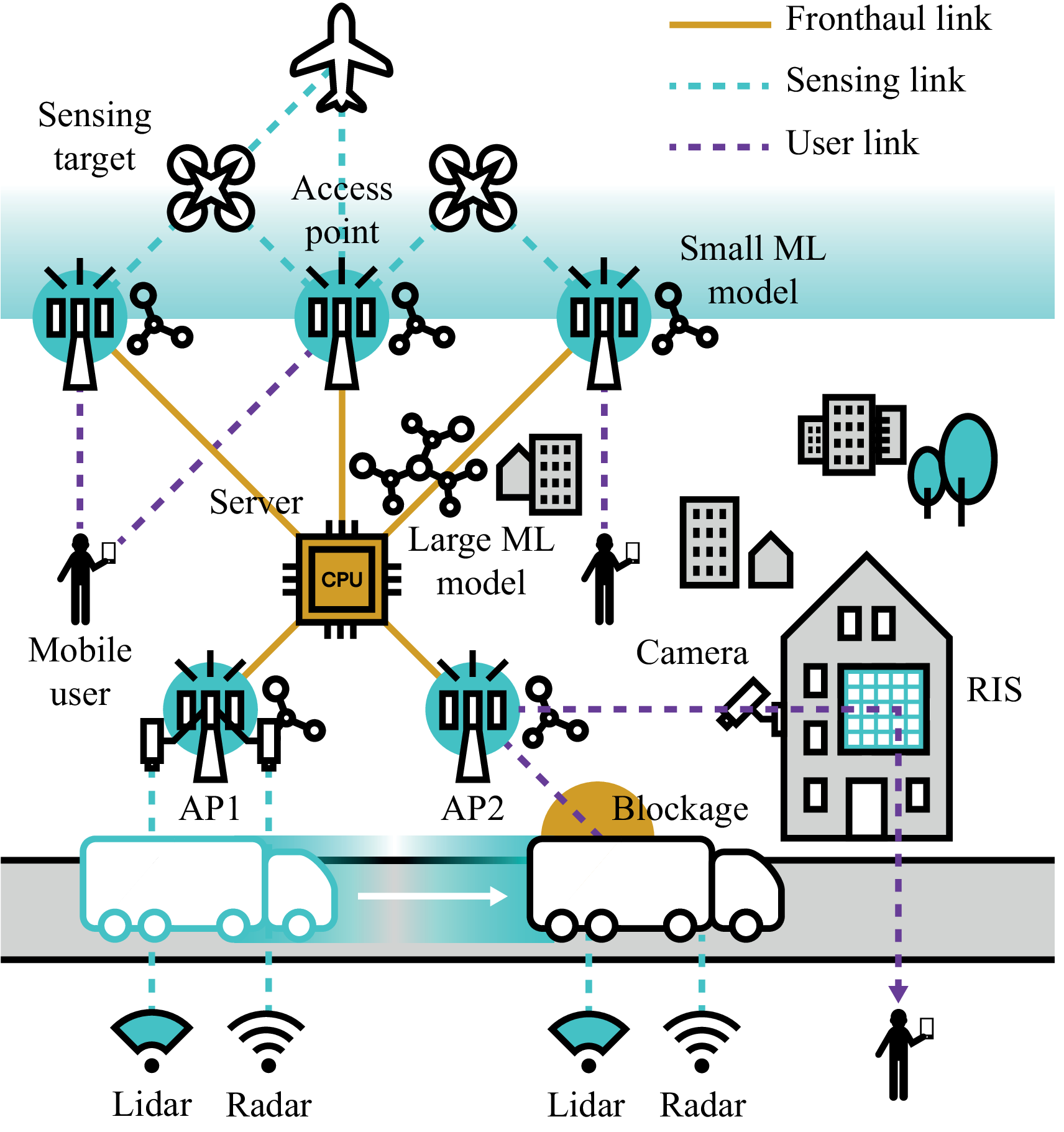}
    \caption{{Collaborative learning in distributed C\&S networks.}}
    \label{fig_netwrok}
\end{figure}

Sixth-generation (6G) wireless networks are envisioned as a transformative platform that integrates communications, sensing, and computing to enable intelligent applications such as smart cities, autonomous transportation, and immersive environments~\cite{giordani2020toward,IMT2030}. Distributed communications and sensing (C\&S) systems, illustrated in Fig.~\ref{fig_netwrok}, are a key architectural enabler of this vision. In such systems, processing capabilities are increasingly shifted from centralized cloud servers to distributed access points (APs) and edge devices~\cite{ngo2017cell}. This shift reduces latency, eases fronthaul congestion, improves scalability, and preserves data privacy, all of which are critical requirements in scenarios with high mobility, massive connectivity, and diverse sensing modalities~\cite{Ali20}.

At the network edge, APs and sensors are expected to carry out complementary C\&S tasks such as environmental perception, user tracking, and beam prediction. Equipping these nodes with artificial intelligence (AI) models tailored to their specific roles and local conditions can greatly enhance responsiveness, lower inference load, and enable real-time, context-aware decision-making. Realizing this vision, however, requires scalable mechanisms for deploying lightweight yet capable AI models across a wide range of heterogeneous edge devices. This need motivates the development of \textit{collaborative learning} frameworks that can propagate intelligence from centralized entities to distributed nodes in a resource-efficient manner.

To realize the full potential of collaborative learning, AI models must meet stringent operational requirements \cite{raviv2023adaptive}. They must support diverse C\&S tasks, such as beam prediction, channel estimation, user localization, and object detection, while maintaining high accuracy. Simultaneously, models must be \textit{lightweight} for real-time inference on edge devices with limited power and memory. Unlike centralized systems, distributed environments are inherently heterogeneous, with varying hardware, sensing modalities, and channel dynamics. This necessitates AI models \textit{adapted to each device} to maximize performance and enable effective collaborative inference \cite{shlezinger2023collaborative}. {Addressing these challenges requires principled methodologies that deliver compact, high-performing, and task-specialized models across the network without imposing excessive computational or communications overhead.}

In this article, we examine {\em knowledge distillation (KD)}~\cite{gou2021knowledge} as a collaborative learning paradigm for distributed C\&S systems. KD has shown promise in wireless tasks such as transceiver design~\cite{Gao21, park2025resource}, channel estimation~\cite{Catak22}, user positioning~\cite{Al24}, and remote sensing~\cite{zhang2021learning}, but mostly as a model compression tool for resource efficiency.  
{In distributed C\&S networks, KD introduces key challenges: coordinating teacher-student models across servers and nodes, defining cooperation levels, and enhancing distributed learning while preserving autonomy. Addressing these challenges requires principled methodologies that deliver compact, high-performing, and task-specialized models across the network without imposing excessive computational or communications overhead.}  
This broader view positions KD as a foundation for collaborative intelligence at the edge, a direction that remains insufficiently explored.

We begin with a concise overview of KD, emphasizing principles that make it suitable for distributed C\&S systems with heterogeneous devices, diverse tasks, and strict edge resource constraints. Building on this, we present KD-based collaborative learning frameworks for future 6G networks, including centralized, decentralized, and semi-centralized strategies, each with distinct trade-offs in complexity, adaptability, communication overhead, and privacy. {Finally, we demonstrate the effectiveness of KD through numerical experiments on a representative C\&S task, namely sensing-assisted beam tracking. The results show that KD-based collaborative learning can significantly reduce model complexity while preserving, and in some cases even improving, C\&S performance at the network edge.}

\begin{figure}
    \centering
    \includegraphics[width=1\linewidth]{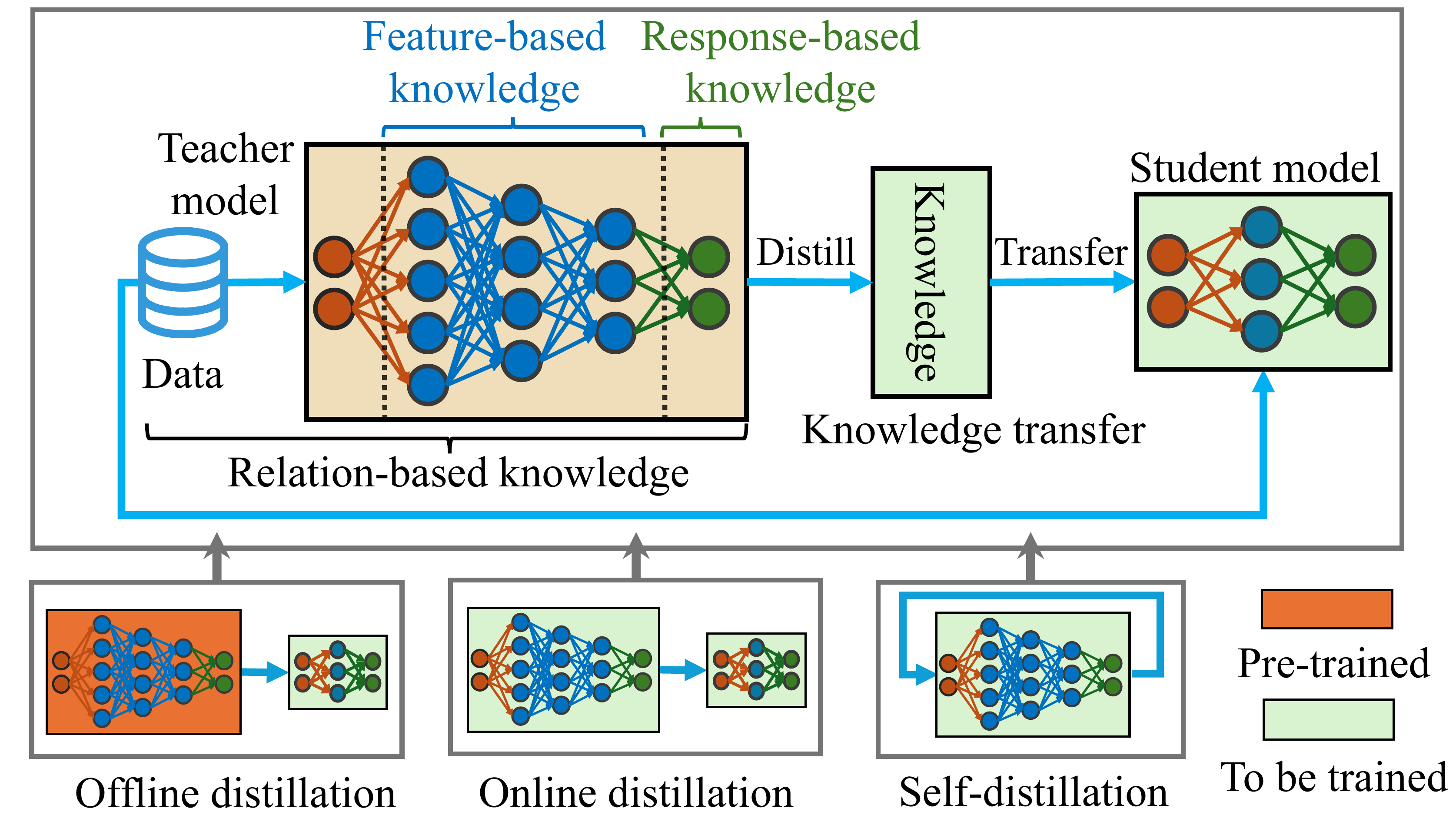}
    \caption{\centering Different types of knowledge and distillation schemes in KD \cite{gou2021knowledge}.}
    \label{fig_KD}
\end{figure}

\section{Knowledge Distillation {for Distributed C\&S}}

We begin with an overview of KD concepts {and highlight its value for distributed C\&S systems}. 
{KD serves both as a model compression technique and a learning framework, enabling smaller models to inherit knowledge from larger ones through different transfer mechanisms and training schemes.}
These foundations motivate how KD can deliver lightweight and efficient AI at distributed network nodes.

\subsection{KD Process}
KD transfers knowledge from a large \textit{teacher} model to a smaller \textit{student}, trained to mimic the teacher under tighter resource constraints. Unlike conventional compression methods such as quantization or pruning~\cite{zhang2021compacting}, KD uses the teacher’s outputs or internal representations as extra supervision. We review two key aspects: the types of {\em knowledge} distilled and the {\em training schemes} employed, as illustrated in Fig.~\ref{fig_KD}, and discuss how these can be adapted for distributed C\&S tasks.

\subsubsection{Types of Knowledge}

A key design choice in KD is the form of knowledge transferred, typically grouped into three categories: \textit{response-based}, \textit{feature-based}, and \textit{relation-based} \cite{gou2021knowledge}.

\textbf{Response-Based Knowledge:} 
This knowledge type refers to the output logits produced by the teacher’s final layer \cite{Gao21, park2025resource, Catak22, Al24}. It is particularly effective for classification tasks, where the soft output distribution can serve as a proxy for the true Bayesian posterior over classes. {Response-based KD allows the student to leverage the teacher’s probabilistic estimates rather than one-hot labels, by minimizing the divergence between their output distributions.} {In distributed C\&S networks, many tasks map naturally to classification problems, making response-based KD useful for beam selection, blockage detection, or signal detection at the edge.}

\textbf{Feature-Based and Relation-Based Knowledge:} {In feature-based KD, the student is trained to align its internal activations with those of the teacher, enabling richer representation learning, particularly in deeper networks.} Relation-based KD further extends this idea by capturing structural relationships such as distances or similarities among samples in the teacher’s latent space \cite{gou2021knowledge, park2025resource}. These knowledge types are not limited to classification and are well suited to a wide range of {C\&S tasks including beamforming, resource allocation, channel estimation, and localization. They also support integration with model-based learning, enabling explainable, efficient, and lightweight machine learning (ML) solutions.}

\subsubsection{Distillation Schemes}

{The second key aspect of KD is the training scheme, which typically follows three paradigms: offline, online, and self-distillation \cite{gou2021knowledge}.}

\textbf{Offline Distillation:}
In this scheme, the teacher is fully trained before student training begins, and the student learns from both ground-truth labels and supervisory signals provided by the teacher \cite{gou2021knowledge, Gao21, park2025resource, Catak22, Al24}. 
{This widely adopted approach assumes access to a pre-trained, high-capacity teacher model.} Such a setting aligns well with distributed C\&S systems, where multiple nodes (e.g., APs) perform similar tasks such as beamforming. A powerful teacher can be trained at a central server with ample resources, and its knowledge distilled into lightweight students for deployment at the APs.

\textbf{Online Distillation:}
Unlike the offline setting, online distillation trains teacher and student models simultaneously, often within a mutual learning framework~\cite{gou2021knowledge}. Each model acts as both teacher and student, enabling knowledge exchange in a co-evolving process. This approach is flexible and does not require a strong pre-trained teacher, though it demands careful synchronization and learning-rate tuning. It is well suited to collaborative distributed C\&S scenarios, where multiple APs or sensors with different data distributions share distilled knowledge in real time to improve tasks such as user tracking and environmental awareness.

\textbf{Self-Distillation:}
A special case of online KD, self-distillation uses a single model as both teacher and student, either across epochs (e.g., earlier vs. later snapshots) or between submodules~\cite{gou2021knowledge}. Although the architecture remains unchanged, self-distillation improves generalization and calibration by reinforcing internal consistency. {Unlike conventional fine-tuning, which updates the model using task-specific supervision, self-distillation incorporates a distillation loss that guides the model using its own previous outputs or intermediate representations.} This approach benefits edge nodes such as APs and sensors under strict privacy constraints, enabling continuous refinement without sharing raw data or relying on external teachers.

{What distinguishes KD is that the student is trained using both direct supervision (e.g., labels or task objectives) and teacher-provided guidance.}
These richer signals strengthen the learning process and improve generalization, particularly for AI models deployed at the edge, where labeled data are scarce, sensing modalities are diverse, and dynamic environments make centralized training impractical.

\subsection{KD for Distributed C\&S}

{This subsection highlights the key properties that make KD particularly valuable for enabling efficient, scalable, and intelligent distributed C\&S deployments at the edge.}

{\textbf{Model Compression with High Task Performance:}}  
A key benefit of offline KD is its ability to compress a high-capacity teacher into a lightweight student while preserving most of the teacher’s performance{~\cite{gou2021knowledge, park2025resource}}. This makes KD especially appealing for distributed C\&S networks, where limited computation, memory, and energy at edge nodes tightly constrain model size. {Other approaches such as quantization and pruning~\cite{zhang2021compacting} also improve resource efficiency, but KD stands out by serving not only as a compression tool but also as a learning framework, both of which are particularly well matched to distributed C\&S systems.}

{\textbf{Specialization and Diversity:}}
Because distillation trains the student on data, it naturally supports tailoring the student model to its local operating context. In distributed networks, edge devices often face heterogeneous environments and perform different C\&S tasks. KD enables each student to be trained on local data, producing models that are not only compact but also personalized and adapted to their device's unique characteristics. This leads to a diverse ensemble of student models across the network~\cite{shlezinger2023collaborative}, each optimized for the specific distributions, dynamics, and objectives of its edge device.

{\textbf{Effective Learning with Limited Data:}}
Another key advantage of KD is its role as a strong regularizer during training. By guiding the student with teacher-generated soft targets or intermediate features, KD enables more stable and data-efficient learning{~\cite{gou2021knowledge}}. This is especially valuable in distributed C\&S settings, where each device may have only limited task-specific data and where collecting extensive labeled datasets is often impractical. KD alleviates data scarcity by transferring the teacher’s inductive biases, improving generalization and reducing overfitting when training on the small datasets typically available at the edge.

{\textbf{Input Reduction and Resource Efficiency:}}  
Beyond model compression, KD also enables input reduction, which is crucial in resource-constrained C\&S applications{~\cite{park2025resource}.} Lowering the dimensionality or frequency of input data greatly reduces energy and processing demands. {This is especially valuable in sensing-aided communications, where multi-modal sensors such as cameras, LiDAR, and radar provide rich context but are power-hungry~\cite{Ali20}}. KD allows students to operate effectively with reduced or single-modality inputs by distilling knowledge from teachers trained on full multi-modal data~\cite{park2025resource}. {It further supports inference from sparse sensing, reducing overhead and latency while preserving task performance.}

\smallskip
The established properties of KD make it a powerful tool for addressing distributed C\&S network challenges. By combining flexibility with resource awareness, KD enables the scalable deployment of diverse AI models across heterogeneous edge devices. The following sections discuss the integration of previously introduced distillation schemes into distributed C\&S via centralized, decentralized, and semi-centralized collaborative learning frameworks, highlighting their respective advantages and limitations in AI-empowered networks.

\section{Integrating KD in Distributed C\&S Networks}

In this section, we explore \textit{collaborative learning} capabilities enabled by integrating KD in distributed C\&S networks. We examine their training dynamics, information exchange mechanisms, and the resulting benefits in terms of scalability, adaptability, and efficiency.

\subsection{Motivation and Rationale}

\begin{figure*}
    \centering
    \includegraphics[width=0.9\linewidth]{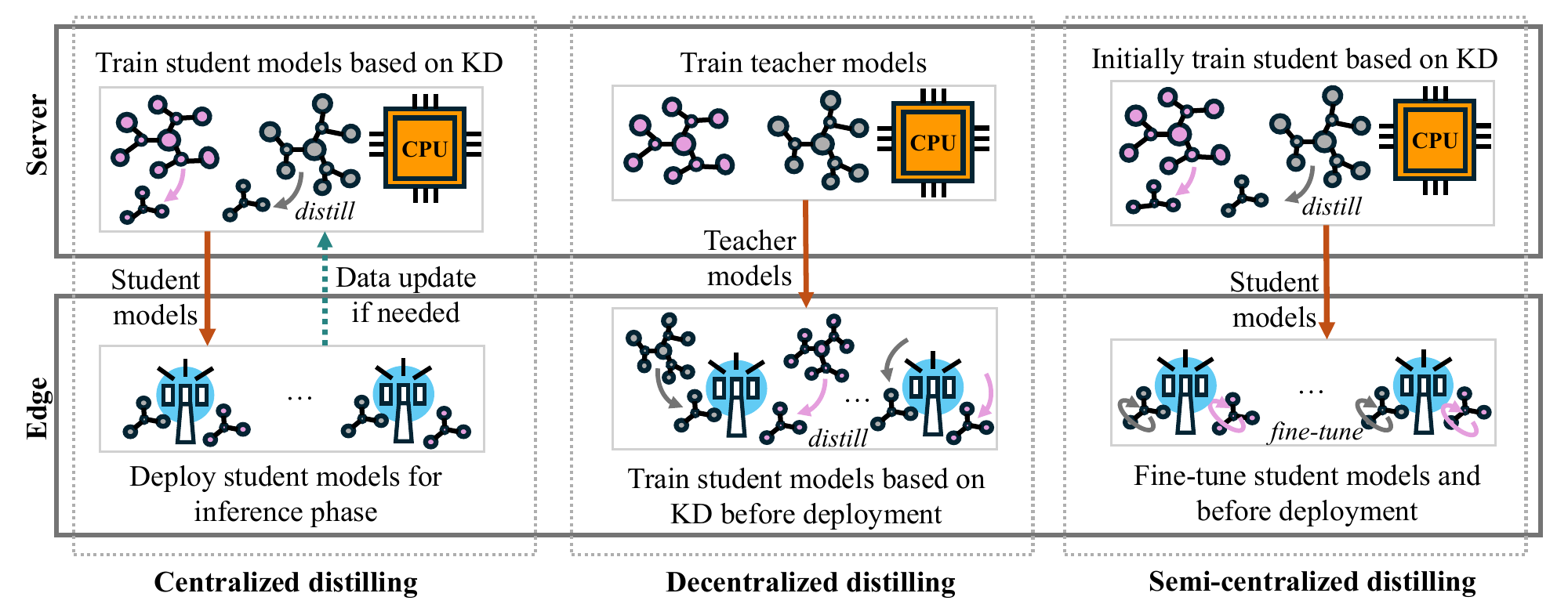}
    \caption{Three knowledge distilling topologies with dynamic deployments of teacher and student models in {collaborative} learning for distributed C\&S networks.}\vspace{-10pt}
    \label{fig_collaborative_learning}
\end{figure*}

In distributed C\&S networks, {lightweight student models are naturally deployed} at nodes such as APs, sensors, and UEs, while a high-capacity teacher resides on a centralized platform (e.g., a server). The centralized teacher, trained on rich and diverse data, {can periodically update students via fronthaul links, enabling efficient knowledge transfer and coordination. This mitigates the limitations of purely local training, where students often underperform due to scarce data and limited resources. KD effectively allows the teacher to guide students, improving their performance despite local constraints}. Furthermore, through online distillation, distributed nodes can collaborate without a pre-trained teacher, exchanging information and intermediate representations to support one another. {These configurations enhance collective intelligence and efficiency in distributed C\&S systems while reducing the burden of individual training, especially in online learning scenarios}.

{While the term {\em collaborative learning} is often associated with federated learning (FL), the two paradigms are designed for different purposes. FL focuses on the centralized training of a shared global model by aggregating updates from distributed edge nodes, typically without exchanging raw data. In contrast, this work considers a KD-based collaborative learning approach aimed at training and deploying multiple edge-specific AI models through teacher-student interactions within distributed C\&S systems. In this setting, the server plays the role of a powerful teacher, while the resulting student models are directly deployed for inference at the edge. Moreover, unlike FL, which commonly enforces uniform model architectures and input/output structures across nodes, KD-based collaborative learning naturally supports heterogeneous models tailored to different devices, sensing modalities, and tasks, making it particularly well suited for distributed C\&S applications.}

\subsection{KD Strategies for Distributed C\&S via Collaborative Learning}

For distributed C\&S systems, as in the common KD setting,  the teacher model is typically trained at a server based on simulations and digital twins. However, the deployment and training of student models can vary depending on system constraints and applications. We identify three primary collaborative learning topologies for integrating KD in such settings, as illustrated in Fig.~\ref{fig_collaborative_learning}:

\textbf{{Centralized Distilling:}} 
In this topology, student models are trained entirely at the server using the available training data and then deployed to edge nodes for inference. This approach leverages the server’s computational resources and minimizes the training burden on edge devices. {Since the student models are lightweight, the communication overhead for deployment is typically low.} However, a key limitation is that the training dataset at the server may become outdated and not accurately reflect the C\&S environments at all local nodes. {Such data mismatch can lead to degraded student performance in dynamic or heterogeneous settings.} Addressing this issue may require frequent dataset updates and model retraining at the server, {which can introduce latency and limit adaptability.}

\textbf{Decentralized Distilling:} 
In this approach, the teacher model is first trained at the server and then transferred to distributed nodes, where it is used to train student models locally. Compared to centralized distilling, this incurs higher communication overhead due to the transfer of the teacher model, but it offers greater flexibility. {In particular, local retraining or fine-tuning allows students to adapt to node-specific data distributions and suboptimal distillation hyperparameters (e.g., temperature or weighting factors in the KD loss function).} This approach also supports data privacy, as local datasets remain at the edge. A potential limitation is that {high-capacity teacher models may still be too resource-intensive to execute at edge devices, even if used only during training.}

\textbf{Semi-Centralized Distilling:} 
This hybrid topology combines the strengths of centralized and decentralized distilling. Student models are first trained at the server and then fine-tuned at edge nodes using local, potentially non-shareable data. This alleviates performance degradation due to data mismatch while keeping communication overhead low. {It also enables local adaptation without deploying the teacher at the edge, but introduces the risk of catastrophic forgetting where the student may partially lose the teacher’s knowledge.}  {To reduce forgetting, a small subset of global or synthetic data can be used for rehearsal, and local loss can be combined with distillation loss to maintain alignment with the teacher’s outputs.} {Effective adaptation still depends on properly chosen distillation parameters during initial training.}

\begin{table*}[t]\small
\centering
\caption{\centering {Comparison of KD topologies in distributed C\&S networks}}
\label{tab:KD_topologies}
\renewcommand{\arraystretch}{1.0}
\setlength{\tabcolsep}{6pt}
\begin{tabular}{p{0.15\linewidth} p{0.25\linewidth} p{0.25\linewidth} p{0.25\linewidth}}
\toprule
\textbf{Aspect} & \textbf{Centralized distillation} & \textbf{Decentralized distillation} & \textbf{Semi-centralized distillation} \\
\midrule
\textbf{Teacher model} & Trained and executed at the server & Trained at the server and transferred to edge for local student training & Trained and executed at the server \\
\midrule
\textbf{Student model} & Trained at the server and deployed to nodes for inference & Trained and deployed locally using the transferred teacher and edge data & Hybrid: initially trained at the server, then fine-tuned at the edge \\
\midrule
\textbf{Data utilization} & Relies solely on centralized datasets & Uses local datasets at each node & Combines global and local datasets for improved generalization \\
\midrule
\textbf{Communications overhead} & Low (only for transferring students); increase if data updates are needed & Moderate (for transferring teacher models from server to edge) & Low (only for transferring student models from server to edge) \\
\midrule
\textbf{Edge computational and memory load} & Minimal (only required for running lightweight student models) & High (to execute teacher models while training students) & Moderate (for fine-tuning student models; teacher models not executed) \\
\midrule
\textbf{Adaptability to local conditions} & Limited (students trained on centralized data) & High (students can be retrained or fine-tuned locally) & High (students can be retrained or fine-tuned locally) \\
\midrule
\textbf{Data privacy} & Low (sending local data to server) & High (no need to share local data) & High (no need to share local data) \\
\midrule
\textbf{Typical use cases} & Stable C\&S environments with low edge dynamics & Highly dynamic or privacy-sensitive systems & Dynamic, privacy-sensitive, and resource-constrained deployments \\
\bottomrule
\end{tabular}
\end{table*}

{Table~\ref{tab:KD_topologies} summarizes the three KD topologies, highlighting their trade-offs in computational load, adaptability, communications overhead, and data privacy.} In the next section, we examine deployment strategies and coordination mechanisms that enable effective collaborative learning in distributed C\&S networks.

\subsection{Model Deployments and Coordination}

{In distributed C\&S systems, different edge AI models do not necessarily perform identical tasks or rely on the same types of input data. Even when targeting a common task, models at different nodes often operate under distinct resource constraints and data availability. For instance, in multi-modal sensing, different AI models may exploit different modalities or combinations thereof depending on the environment and available resources. This heterogeneity motivates flexible KD strategies that differ in teacher deployment, maintenance, and student assignment.}

\textbf{Teacher Models:} 
The server can maintain and train multiple teacher models tailored to the specific tasks and requirements of distributed nodes, as illustrated in Fig.~\ref{fig_collaborative_learning}. Based on node requirements, the server may dynamically select one of the distillation topologies discussed earlier (centralized, decentralized, or semi-centralized) to train and update the student models accordingly. Given that the server typically has access to abundant computational resources and datasets, offline training of multiple teacher models does not impose a significant burden. {Moreover, since teacher models are only invoked during distillation rather than inference, their computational cost does not impact real-time operation.} This flexibility enables scalable and adaptive collaborative learning across heterogeneous C\&S deployments.

{\textbf{Student Models:}} 
As illustrated in Fig.~\ref{fig_collaborative_learning}, edge nodes can host multiple student models, each tailored for different tasks or optimized for specific input conditions. {This allows nodes to adapt model selection to runtime constraints such as sensing quality, energy budget, or computational load.} For example, in multi-modal sensing-aided beam prediction, a node may rely on high-quality camera images during daytime operation, while switching to GPS or radar data at night when visual inputs become less reliable. The previously discussed distillation strategies can be applied to coordinate model updates between the server and distributed nodes. {Although multiple student models are stored locally, only one is executed at inference time, ensuring no increase in runtime complexity and minimal memory overhead due to their lightweight design.}

\begin{figure*}[t]
    \centering
    \includegraphics[width=1\linewidth]{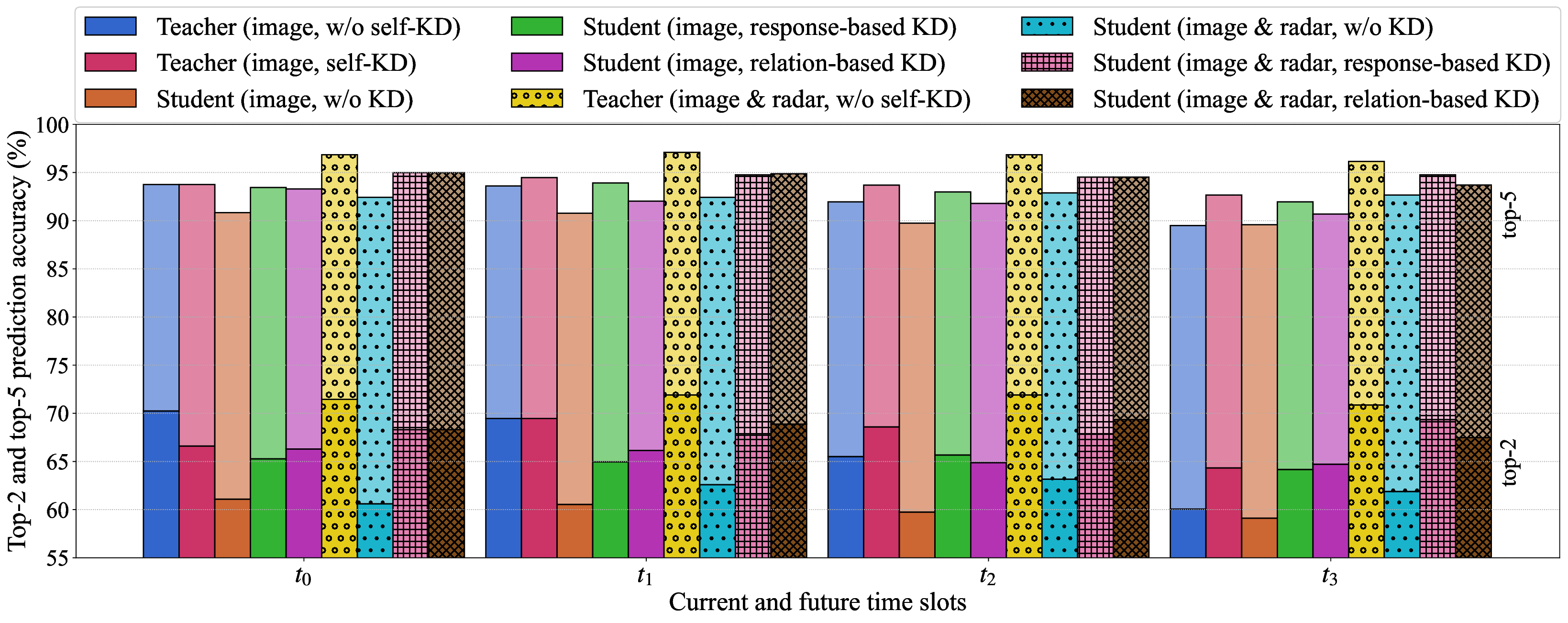}
    \caption{{Beam prediction performance for teacher and student models across current and future time slots, with and without self-KD, response-based KD, and relation-based KD, using mono-modal (image only) or multi-modal (image \& radar) sensing.}}
    \label{fig_topk_acc}
\end{figure*}

\section{Numerical Example}

{In this section, we demonstrate the effectiveness of the response-based KD, relation-based KD, and self-KD strategies discussed earlier, in a distributed C\&S setting. We consider sensing-assisted beam tracking as a representative integrated sensing and communications (ISAC) use case, where the AI model predicts both current and future optimal beams for a vehicle-mounted user based on historical sensing data in a mobile millimeter-wave system. We consider two scenarios, one where the beam is predicted using image data alone, and the other using a combination of image and radar data, both obtained from the real-world DeepSense 6G dataset~\cite{alkhateeb2023deepsense}.\footnote{The source code for this experiment is available at \url{https://github.com/WillysMa/KD-for-sensing}.}}

{We focus on a decentralized distillation scenario in which the teacher is pretrained at the server and deployed at a single edge node to train the student model using the same local dataset. With this setup, the simulations aim to demonstrate the effectiveness of KD in model compression for high-performance sensing-assisted beam tracking. The ability of KD to enable specialization and diversity across edge devices operating in heterogeneous environments and performing different C\&S tasks,  and the implementation of centralized and semi-centralized KD topologies are left for future investigation. Note that the benefit of KD for input and modality reduction {has} already been demonstrated in~\cite{park2025resource}.}

\begin{table*}[t]
\centering
\small
\caption{{Top-2 accuracy (\%) improvement relative to the non-distilled student model, along with model sizes and computational complexities. Student models share the same size and complexity, and the same applies to the teacher models.}}
\label{tab:targetwise_improvement}
\setlength{\tabcolsep}{8pt}
\begin{tabular}{llcccccc}
\toprule
\multirow{2}{*}{\textbf{Category}} & \multirow{2}{*}{\textbf{Model}} & \multicolumn{4}{c}{\textbf{Top-2 accuracy improvement (\%)}} & \multicolumn{2}{c}{{\textbf{Complexity and model size reduction (million)}}} \\
\cmidrule(lr){3-6} \cmidrule(lr){7-8}
& & $t_0$ & $t_1$ & $t_2$ & $t_3$ & {\textbf{No. of parameters}} & {\textbf{No. of FLOPs}} \\
\midrule
\midrule
\multirow{4}{*}{\makecell{Image\\only}} 
  & Teacher (w/o self-KD)  & 9.15 & 8.92 & 5.76 & 0.94 & \multirow{2}{*}{{1.787}} & \multirow{2}{*}{{157.83}} \\
  & Teacher (w/ self-KD)   & 5.52 & 8.92 & 8.84 & 5.21 & & \\
  \cmidrule{2-8}
  & Response-based student & 4.18 & 4.42 & 5.92 & 5.05 & \multirow{2}{*}{{0.095 (18.8$\times$ smaller)}} & \multirow{2}{*}{{92.25 (41.6\% reduction)}} \\
  & Relation-based student & 5.21 & 5.60 & 5.13 & 5.60 & & \\
\midrule
\multirow{3}{*}{\makecell{Image\\ \& radar}} 
  & Teacher (w/o self-KD)  & 10.84 & 9.32 & 8.74 & 8.97 & {2.931} & {179.25} \\
  \cmidrule{2-8}
  & Response-based student & 7.69 & 5.13 & 4.66 & 7.46 & \multirow{2}{*}{{0.106 (27.7$\times$ smaller)}} & \multirow{2}{*}{{42.72 (76.2\% reduction)}} \\
  & Relation-based student & 7.69 & 6.29 & 6.18 & 5.59 & & \\
\bottomrule
\end{tabular}
\end{table*}

{In Fig.~\ref{fig_topk_acc}, we show beam prediction performance in terms of top-2 and top-5 accuracies for the current time slot ($t_0$) and three future slots ($t_1$--$t_3$). Top-$k$ accuracy is the probability that the correct label appears among the model's top $k$ predictions. The detailed performance gains of the distilled students relative to their non-distilled counterparts, as well as the computational efficiency achieved compared to the teacher models, are summarized in Table~\ref{tab:targetwise_improvement}. From Fig.~\ref{fig_topk_acc} and Table~\ref{tab:targetwise_improvement}, we observe:

\begin{itemize}
    \item KD significantly enhances the learning ability of small models, enabling them to outperform their non-distilled counterparts and, in some cases, even the teacher. For example, in the image-only scenario, the relation-based student model achieves $64.72\%$ top-2 accuracy at $t_3$, representing a $5.60\%$ improvement over the non-distilled student model ($59.12\%$). It also surpasses the conventional teacher model without self-KD ($60.06\%$), despite requiring nearly $20$ times fewer parameters.

    \item By leveraging knowledge from the large teacher model with stronger feature extraction capability, training with KD enables the student to process multi-modal sensing data more effectively than independent training. This is evident in the image and radar sensing results. Specifically, at $t_0$, the non-distilled student achieved only $60.61\%$ top-2 accuracy, whereas both the response- and relation-based students reached $68.30\%$. 
    Under the more relaxed top-5 criterion, the distilled student model consistently exhibits gains of over $2\%$ across all time slots.

    \item In terms of model size compression, the student models are nearly $20$ and $30$ times smaller than their corresponding teachers for the image-only and image-plus-radar scenarios, respectively. Despite this substantial compression, the student models still perform close to the much larger teachers. Specifically, for the image-only scenario, the students achieve $64\%$--$66\%$ top-2 accuracy, while the teacher without self-KD ranges between $60\%$ and $70\%$. In the case of image and radar data, the students reach $68\%$--$69\%$, compared to $70\%$--$72\%$ for the teacher models.

    \item Thanks to model compression, the distilled student models achieve significant reductions in computational complexity with minimal loss in accuracy compared to their teachers. Specifically, the students achieve $41.6\%$ and $76.2\%$ reductions in the number of floating-point operations (FLOPs) for the image-only and multi-modal (image and radar) scenarios, respectively. Combined with the reduced model size, this makes the students much more suitable for deployment on edge nodes, {as lower FLOPs and smaller neural networks generally lead to reduced energy and communication costs, although they remain only proxy indicators of these quantities.}   Importantly, these lightweight models still retain substantial performance improvements over non-distilled student models of the same size and computational complexity.

\end{itemize}}

\section{{Challenges} and Open Research Directions}
KD-empowered collaborative learning effectively balances performance against complexity and power consumption. However, practical deployment necessitates overcoming several challenges that expose current limitations and define future research avenues. Key issues include loss function design, student model autonomy, privacy, and standardization, as discussed next.

\subsection{Optimized Loss Functions}
The distillation loss function is central to KD, as it governs how knowledge is transferred from teacher to student. While standard formulations exist for approaches such as response- and relation-based KD~\cite{gou2021knowledge}, refinements are often needed for domain- and task-specific applications. In beam tracking with input reduction, for example, both teacher and student predict current and future beam directions. Here, the loss must be adaptive, reflecting the data availability and prediction horizon: the student may depend more on the teacher for far-future beams, while near-future beams can be effectively learned from labeled data.

\subsection{Enhancing Student Model's Autonomy}
To reduce reliance on teacher models, KD and collaborative learning can be integrated with advanced techniques that enhance the autonomy of student models. {One promising direction is the incorporation of meta-learning, which enables distributed nodes to learn from diverse tasks or datasets. Specifically, the teacher model trained at the server can act as a meta-learner, while distributed student models are trained with a combined loss integrating the task-specific loss, the KD loss from the teacher, and a meta-learning term. This equips student models to retain teacher knowledge while generalizing across varying conditions, such as different channel distributions and hardware impairments. Consequently, student models at edge nodes can quickly adapt to new environments with minimal retraining, reducing the need for frequent teacher execution in decentralized distillation and minimizing the overhead of student retraining and redeployment in centralized distillation.}

{\subsection{Privacy and Security Challenges}

In distributed C\&S scenarios, exchanging models or distilled knowledge between servers and edge nodes creates significant privacy and security risks. A key concern is \emph{information leakage}, where sensitive attributes or sensing characteristics may be inferred from teacher or student models, distilled logits, or intermediate representations through attacks such as model inversion or membership inference. Thus, keeping raw data local does not, by itself, ensure privacy. Another major threat is \emph{model poisoning} or \emph{backdoor attacks}, in which compromised nodes or servers inject malicious models or distillation signals to corrupt student models and manipulate inference. These risks are particularly pronounced in decentralized distillation, where teacher models or knowledge representations are shared, and they also arise in centralized settings when edge nodes upload local data to update teacher models. These risks can be mitigated by sharing only encoded or task-relevant representations and by integrating privacy-preserving techniques into the distillation process. Furthermore, secure channels and encryption help protect model exchanges from eavesdropping, while model integrity checks and anomaly detection assist in identifying poisoned or tampered models before deployment.
}

\subsection{Standardization of KD-Based Collaborative Learning in 6G}

Integrating KD into distributed C\&S systems offers both challenges and opportunities for standardization. Collaborative learning requires frequent exchange of models, distilled outputs, and control signals, necessitating standardized formats for model representation, knowledge transfer, and synchronization. These requirements align with the 6G vision of native AI, emphasizing distributed, adaptive, and intelligent functionality. KD supports this vision by enabling lightweight, locally specialized models that utilize global knowledge without centralizing raw data. Standardizing KD-related mechanisms, such as teacher-student update protocols, model compression interfaces, and secure coordination, will be essential for ensuring interoperability, scalability, and robustness in future 6G networks, in line with ongoing 3GPP efforts to standardize AI/ML model transfer and deployment~\cite{3GPP_AIML2025}.

\section{Conclusion}
In this work, we demonstrated that KD and collaborative learning are key enablers for the efficient and scalable deployment of lightweight AI models in distributed C\&S networks. KD supports edge deployments through model compression, specialization, performance gains, and reduced input requirements, while collaborative learning enables its use in distributed settings via centralized, distributed, and semi-centralized distillation strategies. The dynamic deployment of teacher and student models further allows adaptation to diverse tasks and input conditions. {Numerical results for the decentralized distillation scenario show that} KD-empowered collaborative learning preserves or improves beam tracking accuracy while substantially reducing model complexity, highlighting the practicality of compact, high-performing AI models in distributed environments.

\bibliographystyle{IEEEtran}
\bibliography{IEEEabrv, references}

\section*{Authors' Bio:}

\textbf{Nhan Thanh Nguyen} (nhan.nguyen@oulu.fi) is an Assistant Professor at 
University of Oulu, Finland.

\textbf{Mengyuan Ma} (mengyuan.ma@oulu.fi) is a Postdoctoral Researcher at University of Oulu, Finland.

\textbf{Nir Shlezinger} (nirshl@bgu.ac.il) is an Assistant Professor in School of
Electrical and Computer Engineering in Ben-Gurion University, Israel.

\textbf{Junil Choi} (junil@kaist.ac.kr) is a tenured Associate Professor at School of Electrical Engineering, Korea Advanced Institute of Science and Technology (KAIST), South Korea. 

\textbf{Yonina C. Eldar} (yonina.eldar@weizmann.ac.il) is a Professor in Department of Math and Computer Science, Weizmann Institute of Science, Israel.

\textbf{A. Lee Swindlehurst} (swindle@uci.edu) is a {Distinguished} Professor in the Department of Electrical Engineering \& Computer Science, University of California, Irvine, CA, USA.

\textbf{Markku Juntti }(markku.juntti@oulu.fi) is a Professor
at Centre for Wireless Communications, University of Oulu, Finland.

\end{document}